\documentclass[aps,preprint,nofootinbib,floatfix,superscriptaddress]{revtex4}
\usepackage{epsfig}
\usepackage{graphicx}
\usepackage{multirow}
\usepackage{latexsym,amsbsy,amsmath}
\usepackage{stackrel}
\def\llcol#1#2{\tilde{\lambda}_{#1}.\tilde{\lambda}_{#2}}
\begin{document}
\title{Hidden-flavor pentaquarks}
\author{H.~Garcilazo} 
\email{hgarcilazos@ipn.mx} 
\affiliation{Escuela Superior de F\' \i sica y Matem\'aticas, \\ 
Instituto Polit\'ecnico Nacional, Edificio 9, 
07738 Mexico D.F., Mexico} 
\author{A.~Valcarce} 
\email{valcarce@usal.es} 
\affiliation{Departamento de F\'\i sica Fundamental,\\ 
Universidad de Salamanca, E-37008 Salamanca, Spain}
\date{\today} 

\begin{abstract}
We have recently studied hidden-charm pentaquarks, $c\bar c qqq$,
using dynamical correlations between the heavy quarks arising 
from the Coulomb-like nature of the short-range interaction. 
A pattern was obtained that compares well with the experimental
data. We extend our description to other flavor sectors
which can be framed within the same type of structures 
discussed in the original paper. A detailed 
comparison is made with other results in the literature
and with experimental data. Predictions will be a useful 
tool to discriminate between different models of 
multiquark system dynamics.
\end{abstract}
\maketitle

\section{Introduction}
\label{secI}
The recent findings in the heavy-hadron spectra have become both 
a theoretical challenge and a suitable test bench for trying to understand 
QCD realizations in the non-perturbative regime~\cite{Che16,Bri16,
Ric16,Hos16,Che17,Leb17,Ali17,Esp17,Guo18,Ols18,Kar18,Bra20}.
It appears to be undeniable that more complex quark structures 
allowed by QCD than the simplest quark-antiquark 
(meson) or three-quark (baryon) clusters proposed 
by Gell-Mann~\cite{Gel64}, the so-called multiquarks, 
are being found in the heavy-hadron experimental data~\cite{Aai21,Aaj21}.

For multiquark states with manifestly exotic quantum
numbers, as it is the case of doubly-heavy tetraquarks, both lattice QCD 
approaches~\cite{Hug18,Hud20} and constituent models~\cite{Sil93,Ric18}
predict a very small number of states restricted to very specific
configurations. To give the big picture of multiquark states with ordinary 
quantum numbers it has been suggested the possibility of 
the existence of correlations between the constituents~\cite{Mai15,Gir19,Ali19,Shi21,Wul17}.

In a recent paper~\cite{Gar22}, we have explored
a theoretical scenario where the dynamics of a multiquark system
remains marked by correlations between
heavy flavors dictated by QCD~\cite{Jaf05}, that 
turn the five-body problem into a more tractable three-body problem.
The most suitable system for developing and testing our model
was the hidden-charm pentaquarks, $c\bar c qqq$. We obtained a pattern that compares
well with the experimental data available in this sector.
This work is a natural extension of the analysis performed in Ref.~\cite{Gar22}
to study hidden-flavor pentaquarks which present
the same type of structure in other flavor sectors and, therefore, 
can benefit from the correlations induced by the color-Coulomb potential. 
In particular, we address $b \bar b qqq$ states and also
hidden-charm and hidden-bottom pentaquarks with a strange
quark, $c\bar c qqs$ and $b\bar b qqs$.

The structure of the paper is the following. In the next section
we briefly review the model: the interacting
potential between the quarks, the Hilbert
space arising from the correlations between the heavy flavors,
and the solution of 
the Faddeev equations for the bound state three-body 
problem. In Sec.~\ref{secIII} we present and discuss our 
results. Finally, our 
conclusions are summarized in Sec.~\ref{secIV}.

\section{The model}
\label{secII}
\subsection{Pentaquarks wave function}
\label{subsecII.I}
We study hidden-flavor pentaquarks $Q\bar Q qqq'$, with $Q=b$ or $c$,
$q=u$ or $d$, and $q'=u$ $d$ or $s$. Models based on the attractive 
character of a $qq$ pair in a color-${\bf \bar 3}$ state have been widely
explored in the literature~\cite{Mai15,Gir19,Ali19,Shi21}.
If a $Qq$ color-${\bf \bar 3}$ diquark has a binding proportional to
$m_q$, in the same units a $Q\bar Q$ color-${\bf 1}$ 
has a binding proportional to $2M_Q$. Thus, the color Coulomb-like interaction 
between the components of a hidden-flavor pentaquark
favors a $Q\bar Q$ color singlet instead of a color octet~\cite{Jaf05},
uniquely determining its color wave function. Antisymetrization constraints
allow to identify the different vectors that contribute to
any $(I,J)$ pentaquark for the lowest lying states, i.e., in the case of a fully 
symmetric radial wave function,
{\begin{equation}
\Psi^{(I,J)}_{\rm Pentaquark} \,\, = \,\, \{{\bf 3}_c, i_1, s_1=1/2 \}_{q} \, \otimes \, 
\{{\bf 1}_c, i_2=0, s_2 \}_{(Q \bar Q)} \, \otimes \, 
\{{\bf \bar 3}_c, i_3=s_3, s_3\}_{(qq)} \, ,
\label{ecu1}
\end{equation}}
where $i_1=1/2$ for $Q\bar Q qqq$ pentaquarks and $i_1=0$ for $Q\bar Q qqs$ states.

Table~\ref{tab1} of Ref.~\cite{Gar22} summarizes the possible value of the quantum
numbers leading to an allowed $(I,J)$ hidden-flavor pentaquark. 
Quark correlations dominating the QCD phenomena~\cite{Jaf05}
hint to the most favorable states that can be observed in nature. First,
the very strong quark-antiquark correlation in the color-, flavor-, and spin-singlet 
channel $\{{\bf 1}_c, {\bf 1}_f, 0_s\}$ which can be viewed as 
the responsible for chiral symmetry breaking. The attractive forces in this channel are so
strong that condenses in the vacuum, breaking $SU(N_f)_L \times SU(N_f)_R$ chiral symmetry.
The next most attractive channel in QCD seems to be the color antitriplet, 
flavor antisymmetric, spin singlet $\{{\bf \bar 3}_c, {\bf \bar 3}_f, 0_s\}$, that would select
the $qq$ configurations most important spectroscopically. Thus, we summarize in Table~\ref{tab1}
those states that contain at least one the most attractive QCD channels,
i.e., a diquark with spin zero. $s_1$ stands for the spin
of the single light quark (with isospin $1/2$ for $u,d$ and $0$ for $s$), $s_2$ denotes the spin of the
heavy quark-antiquark pair (with isospin zero) and finally $s_3$ represents the spin of the
light quark pair (with the restrictions imposed by the Pauli principle such that $s_3=i_3$).
The notation in the last column will later be used to identify the wave 
function of the different pentaquarks.
\begin{table}[t]
\begin{tabular}{cp{0.5cm}cp{1cm}cp{0.35cm}cp{0.35cm}cp{0.35cm}cp{0.35cm}c} \hline\hline
         $I$             && $J$     && $s_1$   && $s_2$    && $s_3$  && Vector \\ \hline
\multirow{4}{*}{$1/2 (0)$}   && $1/2$   && \multirow{4}{*}{$1/2$}   && $0$      && $0$    &&  $v_1$ \\
                         && $1/2$   &&    && $1$      && $0$    &&  $v_2$ \\
                         && $1/2$   &&    && $0$      && $1$    &&  $v_3$ \\
                         && $3/2$   &&    && $1$      && $0$    &&  $w_1$ \\ 												
												\hline
$3/2 (1)$                    && $3/2$   &&$1/2$   && $0$      && $1$    &&  $w_3$ \\
\hline \hline
\end{tabular}
\caption{Quantum numbers of the different components resulting in a
$(I,J)$ $Q\bar Q qqq$ hidden-flavor pentaquark containing one of the most attractive QCD channels,
according to Eq.~(\ref{ecu1}). The numbers in
parenthesis stand for the isospin of $Q\bar Q qqs$ pentaquarks. See text for details.} 
\label{tab1}
\end{table}
\subsection{Quark-quark interaction}
\label{subsecII.II}
To perform exploratory studies of systems with more than 
three-quarks it is of basic importance to work with models that correctly
describe the two- and three-quark problems of which thresholds are made 
of. In Ref.~\cite{Gar22} we have adopted a generic constituent model, containing chromoelectric and 
chromomagnetic contributions, tuned to reproduce the masses of the mesons and baryons 
entering the various vectors, the so-called AL1 model by Semay and Silvestre-Brac~\cite{Sem94}.
It has been widely used in a number of exploratory studies of multiquark systems~\cite{Sil93,Ric18,Jan04,Her20,Ric17,Hiy18,Meg19}. It includes a standard 
Coulomb-plus-linear central potential, supplemented by a smeared version of the chromomagnetic interaction,
\begin{align}
\label{ecu3}
V(r)  & =  -\frac{3}{16}\, \llcol{i}{j}
\left[\lambda\, r - \frac{\kappa}{r}-\Lambda + \frac{V_{SS}(r)}{m_i \, m_j}  \, \vec \sigma_i \cdot \vec\sigma_j\right] \, ,\\ \nonumber \\
V_{SS}  &= \frac{2 \, \pi\, \kappa^\prime}{3\,\pi^{3/2}\, r_0^3} \,\exp\left(- \frac{r^2}{r_0^2}\right) ~,\quad
 r_0 =  A \left(\frac{2 m_i m_j}{m_i+m_j}\right)^{-B} \, , \nonumber
\end{align}
where
$\lambda=$ 0.1653 GeV$^2$, $\Lambda=$ 0.8321 GeV, $\kappa=$ 0.5069, $\kappa^\prime=$ 1.8609,
$A=$ 1.6553 GeV$^{B-1}$, $B=$ 0.2204, $m_u=m_d=$ 0.315 GeV, $m_s=$ 0.577 GeV, $m_c=$ 1.836 GeV and $m_b=$ 5.227 GeV. 
Here, $\llcol{i}{j}$ is a color factor, suitably modified for the quark-antiquark pairs.
Note that the smearing parameter of the spin-spin 
term is adapted to the masses involved in the quark-quark or quark-antiquark pairs. 
The parameters of the AL1 potential are constrained in a 
simultaneous fit of 36 well-established mesons and 53 baryons,
with a remarkable agreement with data, as could be seen in Table~2 of Ref.~\cite{Sem94}.
It is worth to note that although the $\chi^2$ obtained in Ref.~\cite{Sem94} with the AL1 potential
is slightly larger than the one obtained with other models, this is essentially
because a number of resonances with high angular momenta were considered. The AL1 model
is very well suited to study the low-energy hadron spectra~\cite{Sil96}. 
The spin-color algebra of the five-quark system has been worked elsewhere~\cite{Ale11,Ric17}.
\subsection{Faddeev equations}
\label{subsecII.III}
The flavor-independence 
of the interacting potential makes the five-body problem to factorize 
into a three-body problem that can be exactly solved by means of the 
Faddeev equations. We follow the method developed in Ref.~\cite{Gar03},
that it is described in detail in Ref.~\cite{Gar22} for $S$- and 
$P$-wave states. Three-body states in which a particle has a given spin can only
couple to other three-body states in which that particle has the same spin, 
since the spinors corresponding to different eigenvalues are orthogonal.
The same applies for isospin. This leads to a decoupling of the integral 
equations in various sets in which the spin and isospin of each particle 
remains the same. The different sets contributing to each $(I,J)$ state
are listed in Ref.~\cite{Gar22}. 

For $S$-wave states one finally gets,
\begin{equation}
T_{i;IJ}^{I_iS_i}(x_iq_i) = \sum_n P_n(x_i) T_{i;IJ}^{nI_iS_i}(q_i) \, ,
\label{eq11}
\end{equation}
where $T_{i;IJ}^{nI_iS_i}(q_i)$ satisfies the one-dimensional integral equation,
\begin{equation}
T_{i;IJ}^{nI_iS_i}(q_i) = \sum_{j\ne i}\sum_{mI_jS_j}
\int_0^\infty dq_j K_{ij;IJ}^{nI_iS_i;mI_jS_j}(q_i,q_j;E)\;
T_{j;IJ}^{mI_jS_j}(q_j) \, , 
\label{eq12}
\end{equation}
with
\begin{eqnarray}
K_{ij;IJ}^{nI_iS_i;mI_jS_j}(q_i,q_j;E)= &&
\sum_r\tau_{i;I_iS_i}^{nr}(E-q_i^2/2\nu_i)
\frac{q_j^2}{2}
\nonumber \\ &&
\times\int_{-1}^1 d{\rm cos}\theta\;
h_{ij;IJ}^{I_iS_i;I_jS_j}
\frac{P_r(x_i^\prime)P_m(x_j)} 
{E-p_j^2/2\eta_j-q_j^2/2\nu_j} \, .
\label{eq13} 
\end{eqnarray}
The three amplitudes $T_{1;IJ}^{rI_1S_1}(q_1)$, $T_{2;IJ}^{mI_2S_2}(q_2)$,
and $T_{3;IJ}^{nI_3S_3}(q_3)$ in Eq.~(\ref{eq12}) are coupled together.

In these equations $\tau_{i;I_iS_i}^{nr}(e)$ are expansion coefficients given 
in terms of Legendre polynomials and the two-body amplitudes $t_{i;I_iS_i}$,
\begin{equation}
\tau_{i;I_iS_i}^{nr}(e)= \frac{2n+1}{2}\frac{2r+1}{2}\int_{-1}^1dx_i
\int_{-1}^1 dx_i^\prime\; P_n(x_i) 
t_{i;I_iS_i}(x_i,x_i^\prime;e)P_r(x_i^\prime) \, ,
\label{eq10} 
\end{equation}
$h_{ij;IJ}^{I_iS_i;I_jS_j}$ are the spin--isospin coefficients
\begin{eqnarray}
h_{ij;IJ}^{I_iS_i;I_jS_j}= &&
(-)^{I_j+i_j-I}\sqrt{(2I_i+1)(2I_j+1)}
W(i_ji_kIi_i;I_iI_j)
\nonumber \\ && \times
(-)^{S_j+s_j-J}\sqrt{(2S_i+1)(2S_j+1)}
W(s_js_kJs_i;S_iS_j) \, , 
\label{eq6}
\end{eqnarray}
where $W$ is a Racah coefficient and $i_i$, $I_i$, and $I$ 
($s_i$, $S_i$, and $J$) are the isospins (spins) of particle $i$,
of the pair $jk$, and of the three--body system. $\eta_i$ and $\nu_i$ are the corresponding reduced masses,
\begin{eqnarray}
\eta_i &=& \frac{m_jm_k}{m_j+m_k} \, , \nonumber\\
\nu_i &=& \frac{m_i(m_j+m_k)}{m_i+m_j+m_k} \, ,
\label{eq3}
\end{eqnarray}
$\vec p_i^{\; \prime}$
is the momentum of the pair $jk$ (with $ijk$ an even permutation of
$123$) and $\vec p_j$ is the momentum of the pair
$ki$  which are given by,
\begin{eqnarray}
\vec p_i^{\; \prime} &=& -\vec q_j-\alpha_{ij}\vec q_i \, , \nonumber\\
\vec p_j &=& \vec q_i+\alpha_{ji}\vec q_j \, ,
\label{eq3p}
\end{eqnarray}
where,
\begin{eqnarray}
\alpha_{ij} &=& \frac{\eta_i}{m_k} 
\, , \nonumber\\
\alpha_{ji} &=& \frac{\eta_j}{m_k} 
\, ,
\label{eq3pp}
\end{eqnarray}
so that,
\begin{eqnarray}
p_i^\prime &=& \sqrt{q_j^2+\alpha_{ij}^2q_i^2+2\alpha_{ij}
q_iq_j{\rm cos}\theta} \, , \nonumber \\
p_j &=& \sqrt{q_i^2+\alpha_{ji}^2q_j^2+2\alpha_{ji}
q_iq_j{\rm cos}\theta} \, .
\label{eq5}
\end{eqnarray}
Finally,
\begin{equation}
x_i=\frac{p_i-b}{p_i+b} \, ,
\label{eq7}
\end{equation}
where and $b$ is a scale parameter that has no effect on the solution.

To solve the Faddeev equations for $P$-wave states, we write them symbolically as,
\begin{equation}
T_i=t_ih_{ij}G_0T_j,
\label{eq14}
\end{equation}
that has to be generalized to a matrix equation,
\begin{equation}
\begin{pmatrix}T_i^{01} \\ T_i^{10}\end{pmatrix}
=\begin{pmatrix}t_i^0 \\ t_i^1\end{pmatrix}h_{ij}G_0
\begin{pmatrix}\hat q_i\cdot\hat q_j & \hat q_i\cdot\hat p_j \\
\hat p_i^{\;\prime}\cdot\hat q_j & \hat p_i^{\;\prime}\cdot\hat p_j 
\end{pmatrix}
\begin{pmatrix}T_j^{01} \\ T_j^{10}\end{pmatrix},
\label{eq15}
\end{equation}
where, from Eq.~(\ref{eq3p}), 
\begin{eqnarray}
\hat q_i\cdot\hat q_j &=& {\rm cos}\theta \, ,
\nonumber \\
\hat q_i\cdot\hat p_j &=& \frac{q_i^2+\alpha_{ji}q_iq_j{\rm cos}\theta}{q_ip_j} \, ,
\nonumber \\
\hat p_i^{\;\prime}\cdot\hat q_j &=& \frac{-q_j^2-\alpha_{ij}q_iq_j
{\rm cos}\theta} {p_i^\prime q_j} \, ,
\nonumber \\
\hat p_i^{\;\prime}\cdot\hat p_j &=& \frac{-(1+\alpha_{ij}\alpha_{ji})q_iq_j
{\rm cos}\theta-\alpha_{ji}q_j^2-\alpha_{ij}q_i^2}{p_i^\prime p_j} \, ,
\label{eq16}
\end{eqnarray}
and $p_i^\prime$ and $p_j$ are given by Eq.~(\ref{eq5}).

In general, the two-body amplitudes $t_{i;I_iS_i}$ are obtained
by solving the Lippmann-Schwinger equation,
\begin{equation}
t=V+VG_0t \, ,
\label{eq17}
\end{equation}
where $V$ is the interaction given by Eq.~(\ref{ecu3}).
Due to the reduction from five to three particles, some pairs of
two-body amplitudes are coupled together.
Therefore, in this case one has to solve the coupled equations,
\begin{eqnarray}
t_{11} &=& V_{11}+V_{11}G_0t_{11}+V_{12}G_0t_{21} \, ,
\nonumber\\
t_{21} &=& V_{21}+V_{21}G_0t_{11}+V_{22}G_0t_{21} \, ,
\label{eq19}
\end{eqnarray}
where the diagonal interactions $V_{11}$ and $V_{22}$ show contributions
from the chromoelectric and chromomagnetic terms of the interaction, while
the off-diagonal interactions $V_{12}$ and $V_{21}$ contain only the
contribution of the chromomagnetic part of the interacting potential.
As expected, the confinement and Coulomb terms are the dominant ones such that the
spin-spin term is just a small perturbation. The effect of the
non-diagonal terms is very small and it can be safely neglected.

\section{Results}
\label{secIII}
We have solved the three-body problem for the different $(I,J)$ $Q\bar Q qqq'$ 
states as discussed in Sec.~\ref{secII}. We show in Table~\ref{tab2} 
the binding energy of the most favorable five-quark states that could be 
observed in nature.
\begin{table}[t]
\begin{ruledtabular}
\begin{tabular}{cccccccc} 
$Q$ & $q$   & $q'$  & $v_1$ & $v_2$ & $v_3$ & $w_1$ & $w_3$ \\ \hline
$c$ & $u,d$ & $u,d$ & 7     & 17    & 24    & 12    & 12    \\
$c$ & $u,d$ & $s$   & 133   & 138   & 143   & 134   & 134   \\
$b$ & $u,d$ & $u,d$ & 39    & 41    & 52    & 40    & 40   \\
$b$ & $u,d$ & $s$   & 165   & 167   & 175   & 166   & 166 \\
\end{tabular}
\caption{Binding energy, in MeV, of the different $Q\bar Q qqq'$ pentaquarks.}
\label{tab2}
\end{ruledtabular}
\end{table}

Let us first note the degeneracy existing between $I=1/2$ and $I=3/2$
$Q\bar Q qqq$ $J=3/2$ pentaquarks ($I=0$ and $I=1$ $Q\bar Q qqs$ $J=3/2$ pentaquarks), 
as could have been expected a priori due to the isospin independence of the potential model in Eq.~(\ref{ecu3}),
although the result is not trivial due to the requirements of the Pauli principle. Secondly,
it has been checked that the conclusions dealing with stability or instability of multiquarks survive
variations of the parameters, we have specifically checked that the pattern remains for different
strengths of the spin-spin interaction by modifying the regularization parameter, $r_0$
in Eq.~(\ref{ecu3}).
 
One can determine the general properties of the multiquarks favored by 
the quark correlations dominating the QCD phenomena shown in Table~\ref{tab1}. 
In the charmonium sector, the mass difference between the $Q\bar Q$ $\{{\bf 1}_c, {\bf 1}_f, 1_s\}$
and $\{{\bf 1}_c, {\bf 1}_f, 0_s\}$ correlated states could be assimilated to the
$J/\Psi - \eta_c$ mass difference. Likewise, in the bottomonium sector it 
corresponds to the $\Upsilon - \eta_b$ mass difference.
The mass difference between the $qq$ $\{{\bf \bar 3}_c, {\bf 6}_f, 1_s\}$
and $\{{\bf \bar 3}_c, {\bf \bar 3}_f, 0_s\}$ has been estimated from full lattice QCD simulations
to be in the range of 100$-$200 MeV~\cite{Fra21,Ale06,Gre10}. Thus, we have fixed the effective 
mass difference of the correlated structures considering the following realistic values,
\begin{align}\label{mass}
\Delta M^{c\bar c} = M^{c\bar c}_{\{{\bf 1}_c, {\bf 1}_f, 1_s\}} - M^{c\bar c}_{\{{\bf 1}_c, {\bf 1}_f, 0_s\}} &= 86 \,\, {\rm MeV} \nonumber \, , \\
\Delta M^{b\bar b} = M^{b\bar b}_{\{{\bf 1}_c, {\bf 1}_f, 1_s\}} - M^{b\bar b}_{\{{\bf 1}_c, {\bf 1}_f, 0_s\}} &= 61 \,\, {\rm MeV} \nonumber \, , \\
\Delta M^{qq} = M^{qq}_{\{{\bf \bar 3}_c, {\bf 6}_f, 1_s\}}  - M^{qq}_{\{{\bf \bar 3}_c, {\bf \bar 3}_f, 0_s\}} &= 146 \,\, {\rm MeV} \, .
\end{align}
Then, denoting by $M_0^{Q \bar Q,q}$ the sum of the masses of a spin zero $Q\bar Q$ diquark, a spin zero $qq$ diquark
and a light quark, the mass of the $Q\bar Q qqq$ states in Table~\ref{tab1} would be given by,
\begin{equation}
M_i = M_0^{Q \bar Q,q} - B_i + \Delta M^{Q\bar Q} \, \delta_{s_2,1} + \Delta M^{qq} \, \delta_{s_3,1} \, ,
\end{equation}
where $B_i$ is the binding energy calculated above, see Table~\ref{tab2}. For $Q\bar Q qqs$ states we have a similar
expression,
\begin{equation}
M_i = M_0^{Q \bar Q,s} - B_i + \Delta M^{Q\bar Q} \, \delta_{s_2,1} + \Delta M^{qq} \, \delta_{s_3,1} \, .
\end{equation}

By taking $M_0^{c\bar c,q}=4319$ MeV, one gets the results shown in Table~\ref{tab3}.
\begin{table}[t]
\begin{tabular}{cp{0.3cm}cp{0.3cm}cp{0.3cm}cp{0.3cm}c} 
\hline\hline
Vector  && $(I)J^P$     && $M_{\rm Th}$ (MeV) &&      State          && $M_{\rm Exp}$ (MeV)                \\ \hline
$v_1$   && $(1/2)1/2^-$ && 4312               &&       $P_c(4312)^+$ && $4311.9 \pm 0.7 ^{+6.8}_{-0.6}$     \\ 
$v_2$   && $(1/2)1/2^-$ && 4388               && \multirow{2}{*}{$P_c(4380)^+$} && \multirow{2}{*}{$4380 \pm 8 \pm 29$}             \\
$w_1$   && $(1/2)3/2^-$ && 4393               &&                                &&                       \\
$v_3$   && $(1/2)1/2^-$ && 4441               &&       $P_c(4440)^+$ && $4440.3 \pm 1.3 ^{+4.1}_{-4.7}$     \\ 
$w_3$   && $(3/2)3/2^-$ && 4453               &&       $P_c(4457)^+$ && $4457.3 \pm 0.6 ^{+4.1}_{-1.7}$    \\  
\hline\hline
\end{tabular}
\caption{Properties of the $c\bar c qqq$ pentaquarks in Table~\ref{tab1}.}
\label{tab3}
\end{table}
As can be seen, there is a good agreement between theoretical states 
showing the most important correlations dictated
by the QCD phenomena and the experimental data~\cite{Aai15,Aai19}.
Thus, Table~\ref{tab3} presents a theoretical spin-parity assignment
for the existing hidden-charm pentaquarks. A careful analysis of the
results and a detailed comparison with other approaches in the literature
was performed in Ref.~\cite{Gar22}. 

Now, we can make parameter-free predictions
for the lowest-lying hidden-bottom pentaquarks, 
for which there is still no experimental.
The results are shown in Table~\ref{tab4},
compared to other results available in the literature.
\begin{table}[b]
\begin{tabular}{cp{0.3cm}cp{0.3cm}cp{0.3cm}cp{0.3cm}cp{0.3cm}cp{0.3cm}c} 
\hline\hline
Vector  && $(I)J^P$     && Our model          && Ref.~\cite{Wul17} && Ref.~\cite{Yan19} && Ref.~\cite{Fer19}  \\ \hline
$v_1$   && $(1/2)1/2^-$ && 11062              && 11137.1           && 11080 (11078)     && 10605  \\ 
$v_2$   && $(1/2)1/2^-$ && 11121              && 11148.9           && 11115 (11043)     && 10629  \\
$w_1$   && $(1/2)3/2^-$ && 11122              && 11237.5           && 11124 (11122)     && 10629  \\
$v_3$   && $(1/2)1/2^-$ && 11195              && 11205.0           && $-$               && $-$ \\ 
$w_3$   && $(3/2)3/2^-$ && 11207              && 11370.6           && 11112 (10999)     && $-$  \\  
\hline\hline
\end{tabular}
\caption{Predictions of different models for the mass, in MeV, of the $b\bar b qqq$ pentaquarks in Table~\ref{tab1}.}
\label{tab4}
\end{table}
Ref.~\cite{Wul17} considers a color-magnetic 
interaction to estimate the mass splitting of the different 
states with respect to a reference mass adjusted to 
experimental data. Ref.~\cite{Yan19} makes
use of a chiral quark model and solves the five-body
bound state problem by the Gaussian expansion method.
We quote the results corresponding to the color-singlet 
calculation, which, from a theoretical point of view, would be the closest
to our model, and between parenthesis the
coupled channel calculation including hidden-color channels.
Ref.~\cite{Fer19} presents results of a hadro-quarkonium
model with baryons of $I=1/2$ and two different
chromoelectric polarizability strengths. We show the results
of the model with lower attraction, in which the 
hidden-bottom pentaquarks are located in the 10.6$-$10.9 GeV energy
region~\footnote{The model with higher attraction, derived by considering
charmonia as a pure Coulombic system, predicts the lowest-lying hidden-bottom
pentaquarks in the 10.4$-$10.7 GeV energy region.}. Positive parity states have 
smaller binding energies and appear about 150 MeV above the negative parity states.
In the first two references, which use quark degrees of freedom,
there is a rich spectra of pentaquarks with different isospins, $I=1/2$ and 
$3/2$, and spins, $J^P=1/2^-$, $3/2^-$ and $5/2^-$. We only show
the lowest lying states to compare with those obtained with 
our model. It is interesting to note that most of the 
hidden-bottom pentaquarks predicted by quark substructure models 
are in the same energy region, 11.0$-$11.2 GeV. The hadro-quarkonium 
model finds more deeply bound states. These differences support 
a future experimental effort to look for hidden-bottom pentaquarks
in this energy region, what would be a clear signal to discriminate 
between the different dynamics that may drive to hidden-bottom 
pentaquarks.

Very recently the LHCb Collaboration announced the observation of a new strange pentaquark
$P_{\Psi s}^\Lambda (4338)$ in the decay $B^- \to J/\Psi \Lambda \bar p$ as a resonance
in the $J/\Psi \Lambda$ invariant mass distribution~\cite{Che22}. It has a mass
of $4338.2 \pm 0.7 $ and a minimal quark content $c\bar c uds$. 
One can take advantage of the recent discovery of this hidden-charm pentaquark
with strangeness to tune the free parameter in the strange sector,
by taking $M_0^{c\bar c,s}=4471$ MeV. Thus, one gets the results shown in Table~\ref{tab5}.
\begin{table}[b]
\begin{tabular}{cp{0.3cm}cp{0.3cm}cp{0.3cm}cp{0.3cm}cp{0.3cm}cp{0.3cm}c} 
\hline\hline
Vector  && $(I)J^P$   &&   $M_{\rm Th}$ (MeV) &&  State                      && $M_{\rm Exp}$ (MeV)              &&  $M^\dagger_{\rm Exp}$ (MeV)  \\ \hline
$v_1$   && $(0)1/2^-$ && 4338                 &&  $P_{\Psi s}^\Lambda(4338)$ && $4338.2 \pm 0.7 $                &&  $4338 \pm 0.7$        \\ 
$v_2$   && $(0)1/2^-$ && 4419                 &&                             &&                                  &&                     \\
$w_1$   && $(0)3/2^-$ && 4423                 &&                             &&                                  &&               \\
$v_3$   && $(0)1/2^-$ && 4474                 &&  $P_{cs}(4459)$             && $4458.8 \pm 2.9 ^{+4.7}_{-1.1}$  &&  $4454.9 \pm 2.7$      \\ 
$w_3$   && $(1)3/2^-$ && 4483                 &&                             &&                                  &&  $4467.8 \pm 3.7$      \\  
\hline\hline
\end{tabular}
\caption{Properties of the $c\bar c qqs$ pentaquarks in Table~\ref{tab1}.}
\label{tab5}
\end{table}

Moreover, the LHCb Collaboration has reported evidence of a structure in the $J/\Psi \Lambda$ invariant 
mass distribution obtained from an amplitude analysis of
$\Xi^-_b \to J/\Psi \Lambda K^-$ decays~\cite{Aak21}. The observed structure,
with mass $4458.8 \pm 2.9 ^{+4.7}_{-1.1}$ MeV, is consistent with being due to a 
charmonium pentaquark with strangeness, i.e., minimal quark content
$c\bar c uds$. These two states are collected in the 
penultimate column, $M_{\rm Exp}$, of Table~\ref{tab5}. However,
the structure observed in Ref.~\cite{Aak21} is also consistent with being due to two resonances, with masses 
$4454.9 \pm 2.7$ MeV and $4467.8 \pm 3.7$ MeV. This experimental situation is
reflected in the last column, $M^\dagger_{\rm Exp}$, of Table~\ref{tab5}. The existence of two states
resembles the situation found in the nonstrange sector with the $P_c(4440)^+$ and the 
$P_c(4457)^+$~\cite{Kar21}, two resonances predicted as $J=1/2$ and $3/2$ states, 
but different isospin in our model, see Table~\ref{tab3}. 

As can be seen, there is a good agreement between theoretical states 
showing the most important correlations dictated
by the QCD phenomena and the experimental data~\cite{Aak21,Che22}.
Therefore, Table~\ref{tab5} presents a theoretical 
spin-parity assignment for the existing strange hidden-charm pentaquarks deduced from
our model together with the prediction of a few new states in the same energy region.
\begin{table}[b]
\begin{tabular}{cp{0.3cm}cp{0.3cm}cp{0.3cm}cp{0.3cm}cp{0.3cm}cp{0.3cm}c} 
\hline\hline
Vector  && $(I)J^P$   &&   Our model          && Ref.~\cite{Wul17} && Ref.~\cite{Wan20}      && Ref.~\cite{Hup22}  && Ref.~\cite{Fer20} \\ \hline
$v_1$   && $(0)1/2^-$ && 4338                 && 4362.3            && $4319.4^{+2.8}_{-3.0}$ && 4330               && 4474        \\ 
$v_2$   && $(0)1/2^-$ && 4419                 && 4548.2            && $4456.9^{+3.2}_{-3.3}$ && 4475               && 4522            \\
$w_1$   && $(0)3/2^-$ && 4423                 && 4556.1            && $4423.7^{+6.4}_{-6.8}$ && 4440               && 4522          \\
$v_3$   && $(0)1/2^-$ && 4474                 && 4571.4            && $4463.0^{+2.8}_{-3.0}$ && 4476               && $-$          \\ 
$w_3$   && $(1)3/2^-$ && 4483                 && 4846.4            && $-$                    && $-$                && $-$      \\  
\hline\hline
\end{tabular}
\caption{Predictions of different models for the mass, in MeV, of the $c\bar c qqs$ pentaquarks in Table~\ref{tab1}.}
\label{tab5b}
\end{table}

In Table~\ref{tab5b} we compare our results with others available in the literature.
The perturbative color-magnetic calculation of Ref.~\cite{Wul17} predicts large splittings
among the lowest lying states. The chiral effective field theory potentials of Ref.~\cite{Wan20}
are closer to the results of our model, with the lowest lying states in the 4.3$-$4.4 GeV
energy region. They do not find bound state solutions for $I=1$ channels. 
Ref.~\cite{Hup22} presents results of a chiral
quark model using a variational method with radial wave functions expanded in terms of 
Gaussians. The model, successful in describing the nonstrange pentaquarks,
predicts very small binding energies in the different baryon-meson channels. They
focus on $I=0$ states and do not show results for $I=1$ systems.
The hadro-quarkonium model of Ref.~\cite{Fer20} obtains the larger masses
for the hidden-charm pentaquarks with strangeness, unlike
non-strange and strange hidden-bottom pentaquarks where they found much smaller
masses that quark based models.

Similarly to the hidden-charm sector, we can now make parameter-free predictions
for the lowest-lying strange hidden-bottom pentaquarks, for which experimental 
data are not yet available. The results are shown in Table~\ref{tab6},
compared to other results in the literature.
\begin{table}[b]
\begin{tabular}{cp{0.3cm}cp{0.3cm}cp{0.3cm}cp{0.3cm}cp{0.3cm}cp{0.3cm}c} 
\hline\hline
Vector  && $(I)J^P$   && Our model          && Ref.~\cite{Wul17} && Ref.~\cite{Fer20}    \\ \hline
$v_1$   && $(0)1/2^-$ && 11088              && 11117.7           && 10671   \\ 
$v_2$   && $(0)1/2^-$ && 11147              && 11183.8           && 10695   \\
$w_1$   && $(0)3/2^-$ && 11148              && 11180.2           && 10695   \\
$v_3$   && $(0)1/2^-$ && 11224              && 11301.2           && $-$     \\ 
$w_3$   && $(1)3/2^-$ && 11233              && 11509.0           && $-$     \\  
\hline\hline
\end{tabular}
\caption{Predictions of different models for the mass, in MeV, of the $b\bar b qqs$ pentaquarks in Table~\ref{tab1}.}
\label{tab6}
\end{table}

The strange hidden-bottom pentaquarks are obtained close to the non-strange case.
They are found to be in the 11.1$-$11.3 GeV energy region. This is the same 
conclusion of the lowest lying states of the perturbative 
chromomagnetic model of Ref.~\cite{Wul17}.
Analogously to the hidden-bottom non-strange case, the hadro-quarkonium model 
of Ref.~\cite{Fer20} reports smaller masses~\footnote{The model with higher 
attraction, derived by considering charmonia as a pure Coulombic system, predicts 
the lowest-lying hidden-bottom pentaquarks in the 10.4 GeV energy region.}.
It also predicts smaller splittings than quark-based model between the lowest-lying states.
Thus, hidden-bottom pentaquarks, either with or without strangeness, seems to be
an adequate tool to discriminate
among different models for the dynamics of multiquark systems.
Models based on the quark substructure indicate that future searches
of nonstrange and strange hidden-bottom pentaquarks
should be carried out in the 11 GeV energy region.

It is worth mentioning that other alternatives have been used for the study
of strange and non-strange hidden-flavor pentaquarks. Among them a 
diquark-triquark model~\cite{Zhu16} has been applied to the same systems studied in this
work. The model exploits the color attractive configurations
of a not pointlike triquark with a light cone distribution amplitude for the 
pentaquark. An effective diquark-triquark Hamiltonian based on a spin-orbital 
interaction is established. The results of this model suggest that the $P_c(4380)^+$ 
could correspond to a $J^P=3/2^-$ state with a mass of 4349 MeV and the state 
originally reported as $P_c(4450)^+$ (see Ref.~\cite{Gar22} for details) would 
correspond to a $J^P=5/2^+$ with a mass of 4453 MeV. The lightest state is a 
$J^P=3/2^-$ hidden-charm pentaquark with a mass of 4085 MeV. $J^P=1/2^-$
states are theoretically discussed but their masses are not shown in the paper.
Following the same strategy strange hidden-charm and hidden-bottom
pentaquarks were studied in Ref.~\cite{Zhu16}. For the sake of completeness we quote the lowest
lying state for each quantum numbers in the different flavor sectors:
hidden-charm $(3/2^-,5/2^-,5/2^+)=(4085,4433,4453)$ MeV;
strange hidden-charm $(3/2^-,5/2^-,5/2^+)=(4314,4624,4682)$ MeV;
hidden-bottom $(3/2^-,5/2^-,5/2^+)=(10723,11045,1146)$ MeV;
strange hidden-bottom $(3/2^-,5/2^-,5/2^+)=(10981,11264,11413)$ MeV.

Preliminary analysis of the experimental data in the hidden-charm sector
suggested the coexistence of negative and positive parity pentaquarks 
in the same energy region~\cite{Aai15}.
We have calculated the mass of the lowest positive parity
state, the first orbital angular momentum excitation of the $v_1$ state.
The technical details have been described in Sec.~\ref{subsecII.III}. 
We chose this state because it is made up of the most strongly correlated structures, 
$Q\bar Q$ $\{{\bf 1}_c, {\bf 1}_f, 0_s\}$ and
$qq$ $\{{\bf \bar 3}_c, {\bf \bar 3}_f, 0_s\}$.
Then, it might have a similar mass to negative parity states 
made up of spin 1 structures.
\begin{table}[t]
\begin{tabular}{cp{0.2cm}cp{0.2cm}cp{0.2cm}|c} 
\hline\hline
$Q$ && $q$   && $q'$  && Mass (MeV) \\ \hline
$c$ && $u,d$ && $u,d$ && 4527    \\
$c$ && $u,d$ && $s$   && 4552   \\
$b$ && $u,d$ && $u,d$ && 11273   \\
$b$ && $u,d$ && $s$   && 11291 \\
\hline\hline
\end{tabular}
\caption{Mass of the first positive parity states, two degenerate states
with quantum numbers $J^P=1/2^+$ and $3/2^+$, for the 
different $Q\bar Q qqq'$ pentaquarks.}
\label{tab7}
\end{table}
As can be seen in Table~\ref{tab7}, the lowest lying positive parity hidden-charm pentaquarks
appear above 4.5 GeV, a mass slightly larger than that of 
the states measured so far. The orbital excitation is 
smaller in the hidden-bottom case such that positive parity
states appear closer to the negative parity ground states.
Similarly, most of the theoretical works prefer to 
assign the lowest lying pentaquarks to negative parity states.
Almost degenerate negative and positive parity states
may occur for hidden-flavor pentaquarks that have been detected 
in the same channel but that were formed by different pairs 
of quarkonium-nucleon states~\cite{Eid16}, one of them radially excited.
The assignment of negative and positive parity states to different parity 
Born-Oppenheimer multiplets has already been suggested as a 
plausible solution in the triquark-diquark 
picture of Ref.~\cite{Gir19}. Nevertheless, this issue remains 
one of the most challenging problems in the pentaquark phenomenology
that should be first confirmed experimentally.
 
Multiquark states would show very different decay patterns regarding 
its internal structure~\cite{Wul17}. The decays of the pentaquarks in
Table~\ref{tab1} into an heavy meson + heavy baryon are 
strongly suppressed since decays into open flavor channels can go 
only via $t$-channel exchange by a heavy meson. Due to the content 
of the pentaquarks states they would follow the decays of quarkonium 
excited states, $\Psi(nS) [\Upsilon(nS)]$ and $\eta_c(nS) [\eta_b(nS)]$. 
Thus, multiquarks containing 
a spin zero heavy quark-antiquark pair: $v_1$, $v_3$ and $w_3$
in Table~\ref{tab1}, would be narrower than those with a spin one 
heavy quark-antiquark pair: $v_2$ and $w_1$ in Table~\ref{tab1}. 
This corresponds nicely with the experimental 
observations up to now. However, besides the contribution to the width 
of the substructures that form each pentaquark, one should also 
consider the width due to the bound nature of the system. At this 
point it is worth to mention that the final width of a resonance 
does not come only determined by its internal content, but there 
are significant corrections due to an interplay between the phase 
space for its decay to the detection channel and its mass with 
respect to the hadrons generating the state~\cite{Gar18}.

Finally, it is worth noting that the correlations 
used do not lead to stable multiquarks for any quark substructure, in the same way
the $NN$ short-range repulsion induced by the one-gluon exchange dynamics is not
universal and disappears for other two-hadron channels~\cite{Oka80}. Thus, for example, the
QCD correlations used in this work would not constraint the color wave function of
pentaquarks with anticharm or beauty, $\bar Q qqqq$. Therefore, such systems would not
present bound states, as recently discussed in Ref.~\cite{Ric19}, due to a non favorable 
interplay between chromoelectric and chromomagnetic effects.

\section{Summary}
\label{secIV}

In brief, we have studied hidden-flavor pentaquarks
which can be framed within the same type of structures 
discussed in our recent work of Ref.~\cite{Gar22}. In particular,
we addressed hidden-flavor pentaquarks $Q\bar Q qqq'$, with $Q=b$ or $c$,
$q=u$ or $d$, and $q'=u$ $d$ or $s$.
The color Coulomb-like nature of the short-range one-gluon exchange
interaction leads to a frozen color wave function of the five-body
system, which allows to reduce the problem to a more tractable three-body problem.
The three-body problem has been exactly solved by means of the Faddeev equations.
The interactions between the constituents are deduced from a generic constituent 
model, the AL1 model, that gives a nice description of the low-energy baryon and meson
spectra. 
 
For the non-strange and strange hidden-charm pentaquarks a 
pattern was obtained that compares well with the experimental
data. The tentative spin-parity assignment of the different 
pentaquarks agrees well with other approaches dedicated to 
study a particular set of states. Under the assumption 
that nature favors multiquarks which are made up of correlated 
substructures dictated by QCD, we have estimated the mass of 
the lowest lying pentaquarks. We have considered realistic values 
for the mass difference of the correlated quark pairs. A good
description of the experimental data has been obtained.

For the non-strange and strange hidden-bottom pentaquarks
we present parameter-free predictions. The splitting between
the lowest-lying states is smaller than in the charm sector,
due to the larger mass of the bottom quark appearing in the
chromomagnetic potential. The hidden-bottom pentaquarks are
found to be in the 11.0$-$11.2 GeV energy region. This is a general
conclusion for model based on the quark substructure. Hadro-quarkonium
models predict smaller masses for hidden-bottom pentaquarks, which 
makes them an excellent test bench for testing the dynamics of 
multiquark systems.

The coexistence of negative and positive parity states in the same energy
region appears to be more likely in the bottom sector.
It is worth noting that the correlations 
used do not lead to stable multiquarks for any quark substructure.
Thus, for example, the QCD correlations used in this work would 
not constraint the color wave function of pentaquarks with anticharm 
or beauty.

Bound states and resonances are usually very sensitive to model 
details and therefore theoretical investigations with different 
phenomenological models are highly desirable. We have tried to minimize the
influence of the interacting potential by using a standard constituent model and
we have explored the consequences of dynamical correlations arising from the
Coulomb-like nature of the short-range potential. 
The pattern obtained could be scrutinized against the future experimental results
providing a great opportunity for extending our knowledge to some unreached 
part of the hadron spectra. More such exotic baryons are expected and needed to make
reliable hypotheses on the way the interactions in the system are
shaping the spectra.

\section{Acknowledgments}
This work has been partially funded by COFAA-IPN (M\'exico) and 
by Ministerio de Ciencia e Innovaci\'on and EU FEDER under 
Contracts No. PID2019-105439GB-C22 and RED2018-102572-T.

\end{document}